\pdfoutput=1

\documentclass[]{spie}  
\usepackage[]{graphicx}
\usepackage[usenames,dvipsnames]{color}
\usepackage{rotating}

\title{VLTI status update: a decade of operations and beyond} 

\author{Antoine M\'erand\supit{a}, Roberto Abuter\supit{b}, Emmanuel Aller-Carpentier\supit{b}, Luigi Andolfato\supit{b}, Jaime Alonso\supit{a}, Jean-Philippe Berger\supit{b}, Guillaume Blanchard\supit{a}, Henri Boffin\supit{a}, Pierre Bourget\supit{a}, Paul Bristow\supit{b}, Claudia Cid\supit{a}, Willem-Jan de Wit\supit{a}, Diego del Valle\supit{a}, Fran\c coise Delplancke-Str\"obele\supit{b}, Fr\'ederic Derie\supit{b}, Lorena Faundez\supit{a}, Steve Ertel\supit{a}, Rebekka Grellmann\supit{a},  Philippe Gitton\supit{a}, Andreas Glindemann\supit{b}, Patricia Guajardo\supit{a}, Sylvain Guieu\supit{a}, St\'ephane Guisard\supit{a}, Serge Guniat\supit{b}, Pierre Haguenauer\supit{a}, Christian Herrera\supit{a}, Christian Hummel\supit{b}, Carlos La Fuente\supit{a}, Marcelo Lopez\supit{a}, Pedro Mardones\supit{a}, Sebastien Morel\supit{a}, Andr\'e M\"uller\supit{a}, Isabelle Percheron\supit{b}, Than Phan Duc\supit{b}, Andres Pino\supit{a}, Sebastien Poupar\supit{a}, Eszter Pozna\supit{b}, Andres Ramirez\supit{a}, Sridharan Rengaswamy\supit{a}, Lionel Rivas\supit{a}, Thomas Rivinius\supit{a}, Alex Segovia\supit{a}, Christian Schmid\supit{b}, Markus Sch\"oller\supit{b}, Nicolas Schuhler\supit{a}, Julien Woillez\supit{b}, Markus Wittkowski\supit{b}.
\skiplinehalf
\supit{a} European Southern Observatory, Casilla 19001, Santiago, Chile; 
\skiplinehalf
\supit{b} European Southern Observatory, Karl-Schwarzschild-Str. 2, 85748 Garching, Germany; 
}

\authorinfo{Further author information: Send correspondence to Antoine M\'erand: E-mail: amerand@eso.org}

\begin{document} 
\maketitle 

\begin{abstract}
We present the latest update of the European Southern Observatory's Very Large Telescope interferometer (VLTI). The operations of VLTI have greatly improved in the past years: reduction of the execution time; better offering of telescopes configurations; improvements on AMBER limiting magnitudes; study of polarization effects and control for single mode fibres; fringe tracking real time data, etc. We present some of these improvements and also quantify the operational improvements using a performance metric. We take the opportunity of the first decade of operations to reflect on the VLTI community which is analyzed quantitatively and qualitatively. Finally, we present briefly the preparatory work for the arrival of the second generation instruments GRAVITY and MATISSE.
\end{abstract}


\keywords{Optical Interferometry, Facility, Science Operations Statistics}

\section{INTRODUCTION}
\label{sec:intro}  

The Very Large Telescope Interferometer is a unique facility: not only there are few optical telescope interferometers operated in the world, but only one remains combining large telescopes (8 meter class, known as Unit Telescopes or UT), as well as relocatable smaller telescopes (1.8 meter in diameter, known as te Auxiliary Telescopes, or ATs). For more than 10 years now, the VLTI has been an open time facility, running using the same operational model as the UTs in single dish operation: mixing visitor mode and service mode (or queue mode). 

In the past 10 years, the operations have improved a lot, reaching a high level of efficiency, as we will demonstrate below. On the other hand, the VLTI is about to experience big changes: the second generation of focal instruments (called combiners) will arrive soon, accompanied with a wealth of upgrades on the facility. We will give an overview of the changes to come, but the details can be found in many articles from these proceedings.

\section{STATISTICS ON USAGE AND PERFORMANCE METRICS}

\subsection{ANALYSIS OF THE NIGHT LOGS}

\begin{figure}
   \begin{center}
   \begin{tabular}{c}
   \includegraphics[width=16.5cm]{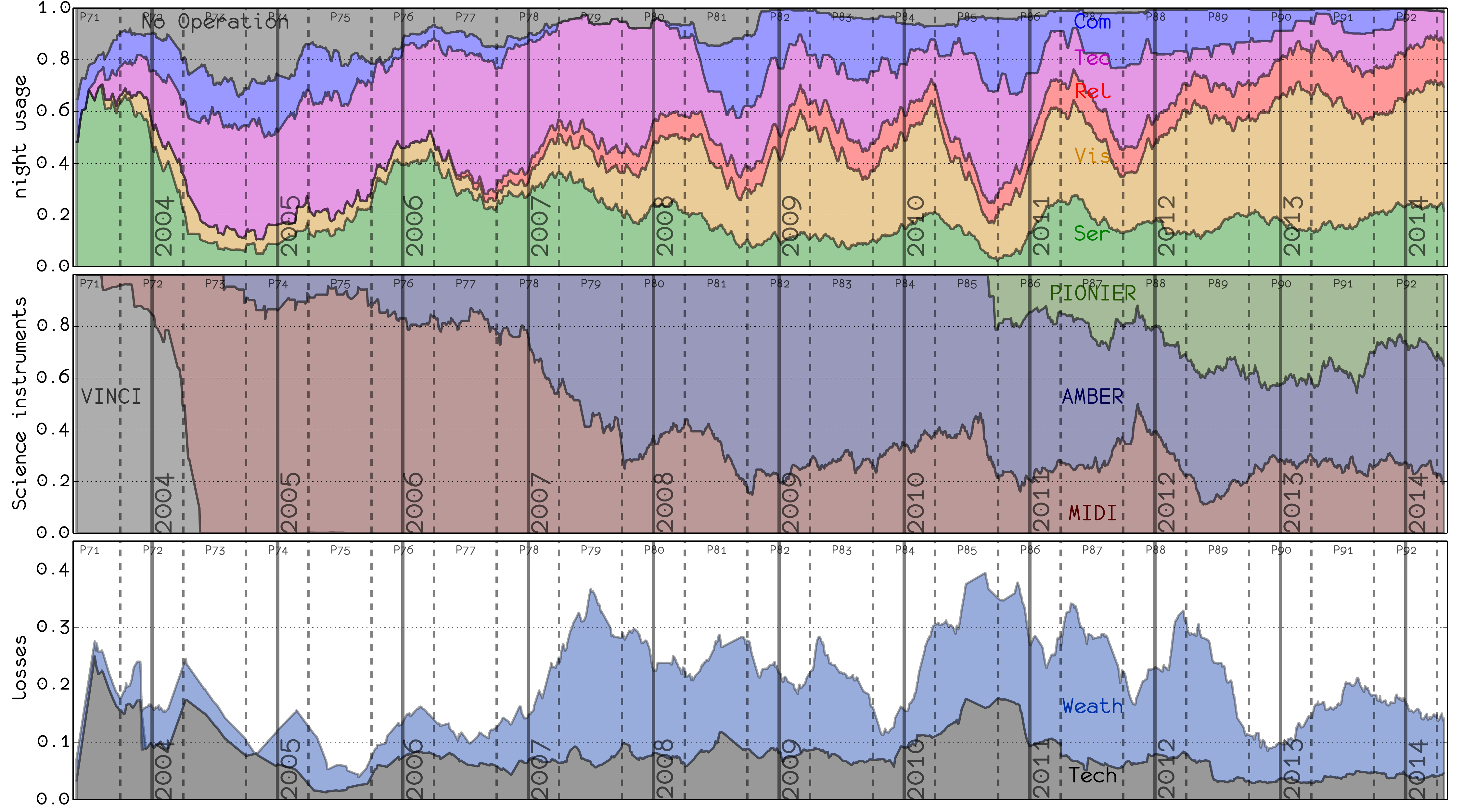}
   \end{tabular}
   \end{center}
   \caption[VLTIusage]{\it \small \label{fig:VLTIusage} Use of VLTI over the last 11 years (2003-2014). \textbf{Upper panel}: Cumulative fractional use, from bottom to top: Service Mode (green), Visitor Mode (yellow), AT after Relocation activities (red), other Technical Activities (purple), Commissioning (blue) and No Operations (gray); \textbf{Middle panel}: cumulative fractional use of Instruments during Science time (Service or Visitor mode), from bottom to top: VINCI (gray), MIDI (brown), AMBER (dark blue) and PIONIER (green); \textbf{Lower panel}: cumulative fraction of science time lost due to technical problems (in gray) or weather (in blue). Note the range of this latest plot is not 0-1 but 0-0.45} 
\end{figure} 

Every night, the activities and losses are logged in order to monitor how we use the VLTI facility. The overview of more than a decade of data can be found on Fig.~\ref{fig:VLTIusage}. The VLTI increasingly improved its efficiency over the years: even though it started operations 10 years ago (MIDI was offered in 2004 and AMBER in 2005), it took several years to routinely operate the facility, i.e. producing scientific data. If we exclude the year 2003, during which the test instrument VINCI was successfully used for commissioning and early science,  we reached 50\% of time dedicated to Science Operations (Service Mode, Visitor Mode or After AT Relocation activities) only in 2007, the rest of the time being used for technical activities or commissioning. As of 2014, about 90\% of the time is dedicated to science operations (service, visitor or after relocation).

Monitoring losses is also an interesting performance indicator. We show on Fig.~\ref{fig:VLTIusage} the two main sources of downtimes: technical problems and weather downtime. The large amount of technical losses (about 10\%) triggered, among other things, the creation of a VLTI System Engineering\cite{vltise} group in 2007 to address the sources of the unreliability of the VLTI and fix them. The amount of technical losses peaked in 2010 at almost 15\%, due to obsolescence related problems of the Delay Lines and the Auxiliary Telescopes. At the same time, in August 2010, occurred a long technical intervention (as seen as a large purple fraction on Fig.~\ref{fig:VLTIusage}, upper panel), which resulted in improvements of the reliability, as seen by the drop in technical loss soon after.

\subsection{WEATHER LOSSES}

\begin{figure}
   \begin{center}
   \begin{tabular}{c}
   \includegraphics[width=16.5cm]{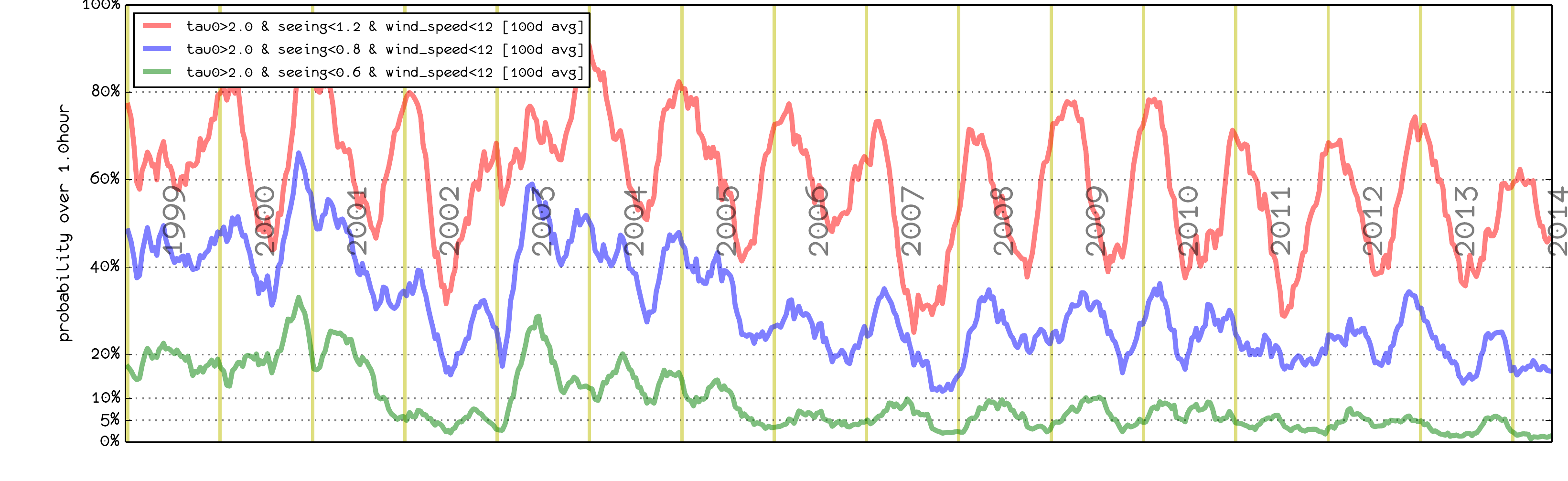}
   \end{tabular}
   \end{center}
   \caption[ASMstats]{\it \small \label{fig:ASMstats} Paranal Observatory weather statistics over the past 15 years, as a function of time. Each curve represents the probability to get different seeing conditions, from top to bottom: 1.2", 0.8" and 0.6", all for wind speed lower than 12m/s and coherence time (in the visible) larger than 2ms. These 3 seeing conditions are the ones offered for the PIs to select from in service mode.} 
\end{figure}

Weather loss is, unfortunately, built-in in the case of seeing limited optical interferometry: bad seeing means bad coupling in the single mode fiber used in the recombining instruments. Bad seeing is also often associated to low coherence time, which forces VLTI instruments to operate at their highest frame rates: high frame rate and low coupling result in low signal-to-noise ratio, hence poor performances. Moreover, the Auxiliary Telescopes (AT) are known to introduce problematic vibrations, degrading the data, in case of wind speed above 12m/s\cite{finito2012}\,. The amount of weather losses is the expected one from the collected environment conditions (seeing, wind speed, cloud coverage, coherence time, etc.) considering the sensitivity of the VLTI instruments. For example, on Fig.~\ref{fig:ASMstats} plotted from ESO Ambient Server\footnote{\texttt{http://archive.eso.org/asm/ambient-server}}, we show the probability of having in a given night the 3 conditions sets, "excellent", "good" and "fair", representative of the conditions offered in service mode (seeing 0.6", 0.8" and 1.2"). The seasonal effects that are seen on this figure are the same that lead to weather loss pattern from the night logs statistics (Fig.~\ref{fig:VLTIusage}). In the future, the ATs will be equipped with adaptive optic modules allowing to overcome part of the weather losses by improving the seeing when the conditions are not optimum (see Sec.~\ref{sec:naomi} below).

\subsection{AUXILIARY TELESCOPES BASELINES CONFIGURATIONS}

\begin{figure}
   \begin{center}
   \includegraphics[width=16.5cm]{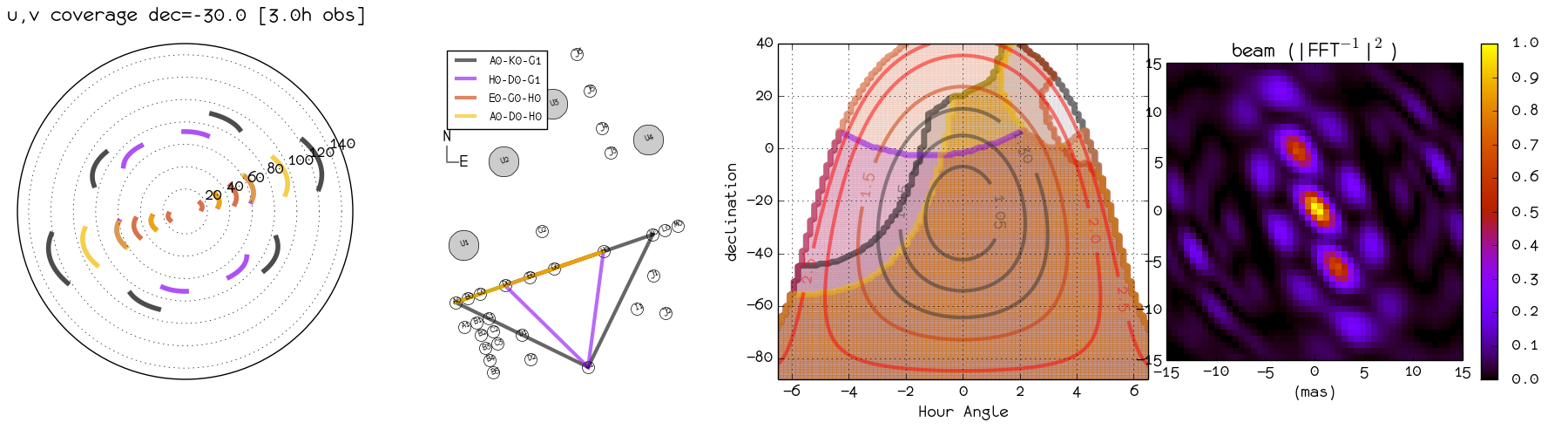} 
   \includegraphics[width=16.5cm]{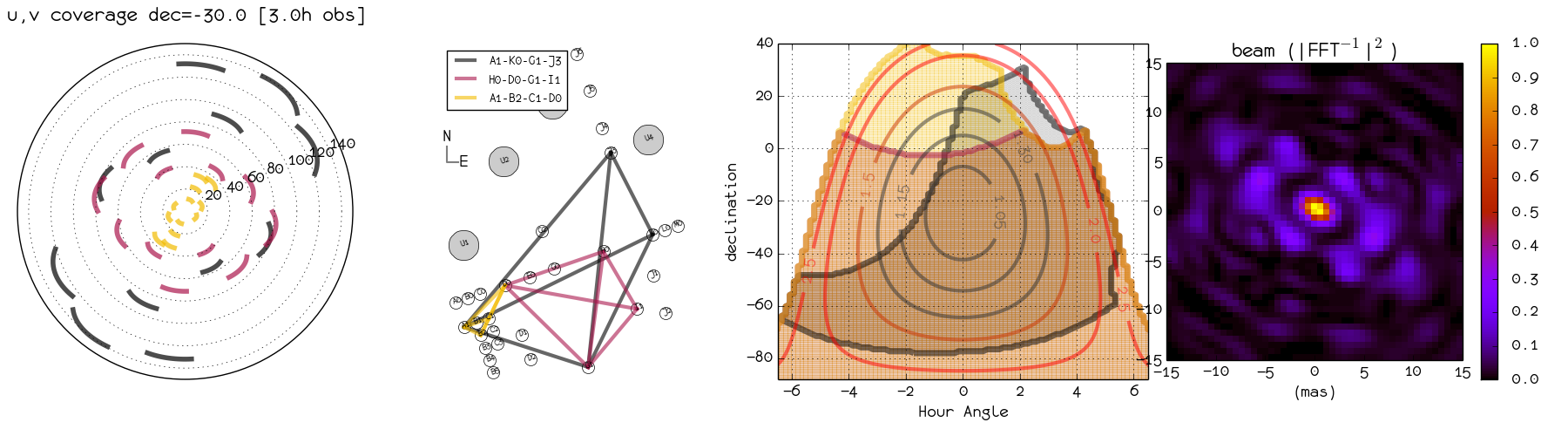}
   \end{center}
   \caption[uvAT]{\it \small \label{fig:uvAT} Offered AT configuration in 2006 (upper panels) and 2012 (lower panels). From left to right: 1) u,v coverage for 3h of observation at declination -30 degrees; 2) baseline map 3) declination/hour angle sky coverage, taking into account UT shadowing and delay lines limits 4) "dirty beam" reconstructed from the u,v map.}
\end{figure} 

Part of the effort to improve the performances of the VLTI came in the form of optimizing the offering of the Auxiliary Telescopes configurations. ATs are movable telescopes, so various spatial frequencies can be sampled using a limited number of telescopes. The offering of AT configurations has to be thought within many constraints: we want to limit the number of configurations for operational reasons; also, each baselines, triplet and quadruplets have limited sky coverage due to the lengths of the delay lines, etc. If we take three different dates at VLTI, we can see how the offering of the configurations evolved:

\begin{itemize}
\item \textbf{2006}: we offered 4 telescopes triplets. On a given night, \textbf{each triplet provides 3 unique u,v points};
\item \textbf{2010}: we offered 3 telescopes quadruplets. Each quadruplet provides, on a given night, 6 possible u,v points. However, the different quadruplets all had 2 telescopes in common, \textbf{so each configuration offers only 4 unique u,v points on average;}
\item \textbf{2012}: since then, we offer 3 telescopes quadruplets. Each quadruplet provides, on a given night, 6 possible u,v points. In the case of these new configurations, the different quadruplets have only 1 telescope in common, \textbf{so each configuration offers 6 unique u,v points.}
\end{itemize} 

\noindent In addition, one can see on Fig.~\ref{fig:uvAT} the AT configurations in 2006 and 2012: the u,v plane is much better sampled in 2012, as can be seen on the rightmost panels (dirty beam). These progresses allowed the VLTI to deliver its first reconstructed images in 2009\cite{2009A&A...496L...1L, 2009A&A...497..195K}\,. Nowadays, imaging is the dominating mode sought by most observing proposals: this implies ever more frequent changes of the AT configuration, known as relocations. Up to 30\% of the time spent on the ATs (as opposed to UTs to other technical activities) is spent doing the verification after the relocations.

\subsection{OPERATION PERFORMANCE METRIC}
\label{sec:perfo}

As can be seen in the previous sections, not one aspect of the operation statistics give a global view on the operational performances. A global performance metric is essential to quantify the progress in operation efficiency. A metric should take into account:
\begin{itemize}
\item The fraction of time spent collecting science data: this should take into account time spent relocating the telescope array and the technical downtimes;
\item The amount of time required to obtain a valid observation. In the case of interferometry, this would be a CAL-SCI sequence: in other word, a calibrated observations of visibilities and/or closure phases;
\item The number of unique u,v points (i.e. projected baseline) accessible in one night. 
\end{itemize}
\noindent We compiled in the following table the efficiency metric for 3 different dates: 2006, 2010 and 2013:

\begin{center}       
\begin{tabular}{|l|c|c|c|c|} 
\hline
  & 2006 & 2010 & 2013  & 2006 $\rightarrow$ 2013\\
\hline
Science Time (sky+reloc) & 40\% & 60\% & 80\% & \textcolor{OliveGreen}{$\times$2}\\
\hline
Technical losses & $\sim$10\% & $\sim$15\% & $\sim$5\% & \textcolor{OliveGreen}{$\div$3}\\
\hline
After relocation nights &  20\% & 25\% & 30\% & \textcolor{BrickRed}{$\times$1.5}\\
\hline
calibrated pointing / hour & 0.5 & 1.0 & 1.5  & \textcolor{OliveGreen}{$\times$3}\\
\hline
unique u,v / config & 3 & 4 & 6 & \textcolor{OliveGreen}{$\times$2}\\
\hline
\hline
\bf effective calibrated pointing / hour & \bf 0.14 & \bf 0.4 & \bf 0.8 & \textcolor{OliveGreen}{\bf $\times$6}\\
\hline
\bf effective calibrated u,v points / hour & \bf 0.4 & \bf 1.6 & \bf 4.8 & \textcolor{OliveGreen}{\bf $\times$12}\\
\hline 
\end{tabular}\end{center}

The last two lines show the global metrics and the quantified big improvements in operations VLTI has witnessed: the "effective / hour" means that, if one takes a random hour of night during this year, we did as many CAL-SCI observation and as many unique u,v points on a given target. The progresses are quantitatively big: \textbf{in 8 years the operational efficiency improved by a factor of 12}. Naively, one would think it transfers into a ten fold increase in publication, but actually the improvements were qualitative: it allows VLTI to produce reconstructed images or constrain complex models with the data, thanks to the wealth of data collected. 

In this table, the only aspect where we lost efficiency is in the increase of after relocation nights activities. These nights are necessary to ensure that the relocatable telescopes are functional after they were moved. We have many movements, since observers often request imaging-like programs, where they want to observe their objects with the 3 different configurations in a short amount of time. Since the configurations have only one telescope in common, each change require to relocate 3 telescopes.

\section{OTHER RECENT IMPROVEMENTS}

\subsection{AMBER SENSITIVITY}
 
The AMBER instrument (3T near-infrared combiner) has seen its sensitivity improved. The guaranteed limiting magnitudes\footnote{\texttt{http://www.eso.org/sci/facilities/paranal/instruments/amber/inst.html}} have been improved thanks to: the replacement of some optics of the spectrograph whose coating was degraded / out of specifications; the realignment of the spectrograph; and the implementation of a polarization control, inspired from PIONIER\cite{niobates}\,, allowing to get rid of the input polarizer, hence recovering (almost) twice as much flux. This improvements provide between 1.5 and 2 magnitudes improvement in term of guaranteed limited magnitudes.

On top of these efforts, members of the AMBER consortium are exploring an experimental "blind" observing mode: individual frames are recorded with very low fringes' signal to noise ratio, and then combined offline. The advantage of the method is, of course, a gain in sensitivity, but at the cost of loss of accuracy and also spectral resolution, since the stacking is done in the 2 diminutional space (fringes' spacing / spectral dispersion) in order to beat the natural fringes' peak frequencies overlap of AMBER in the fringes' spacing dimension. 

\subsection{MIDI+FSU}

For 2 years, we have been offering MIDI (the 2T mid-infrared instrument) with the K-band fringe sensor unit (FSU) of PRIMA as an external fringe tracker. The implementation of the mode is documented in another contribution of this conference. Since we offer the mode, about one third of the MIDI proposals ask specifically for MIDI+FSU. The introduction of the new mode might be the reason for the recent increase of requested time and the increase of new PIs on MIDI (as seen in Fig.~\ref{fig:Community} and discussed in the next section). A dedicated paper of these proceedings describe the implementation of the mode\cite{midifsu}, and a complete description can also be found in a recently published article\cite{midifsuMueller}\,.

\subsection{UT VIBRATIONS}

As presented in 2012\cite{UTvibrations2012}\,, an important amount of work has been done to reduce and control the telescopes vibrations. Two UTs had already reached quite stable levels, but will still benefit from further optimizations. Except for UT2, which has no specifically noisy instruments, the vibration level of the UTs have to be carefully monitored and verified after each intervention on instruments (especially noisy ones) as they could result in important changes.

Between 2008 and 2012, the level of vibrations, in particular in term of optical path difference RMS, went down dramatically: from $\sim$1000nm to $\sim$200nm or less per telescope. This means that, in the K band, the OPD jitter went from $\lambda$/2 to $\lambda$/5. Though still not optimum, this improves a lot the fringe's contrast at the interferometric focus. Recently, even more progresses have been made in the understanding of the UT vibrations phenomena, in particular thanks to a few night of VLTI-UT observations done with all UT instruments turned off\cite{UTvibrations2014}\,. These understandings have yet to lead to substantial decrease of the OPD jitter.

\section{USER COMMUNITY}

\subsection{REQUESTED TIME}

\begin{figure}
   \begin{center}
   \includegraphics[height=9.1cm]{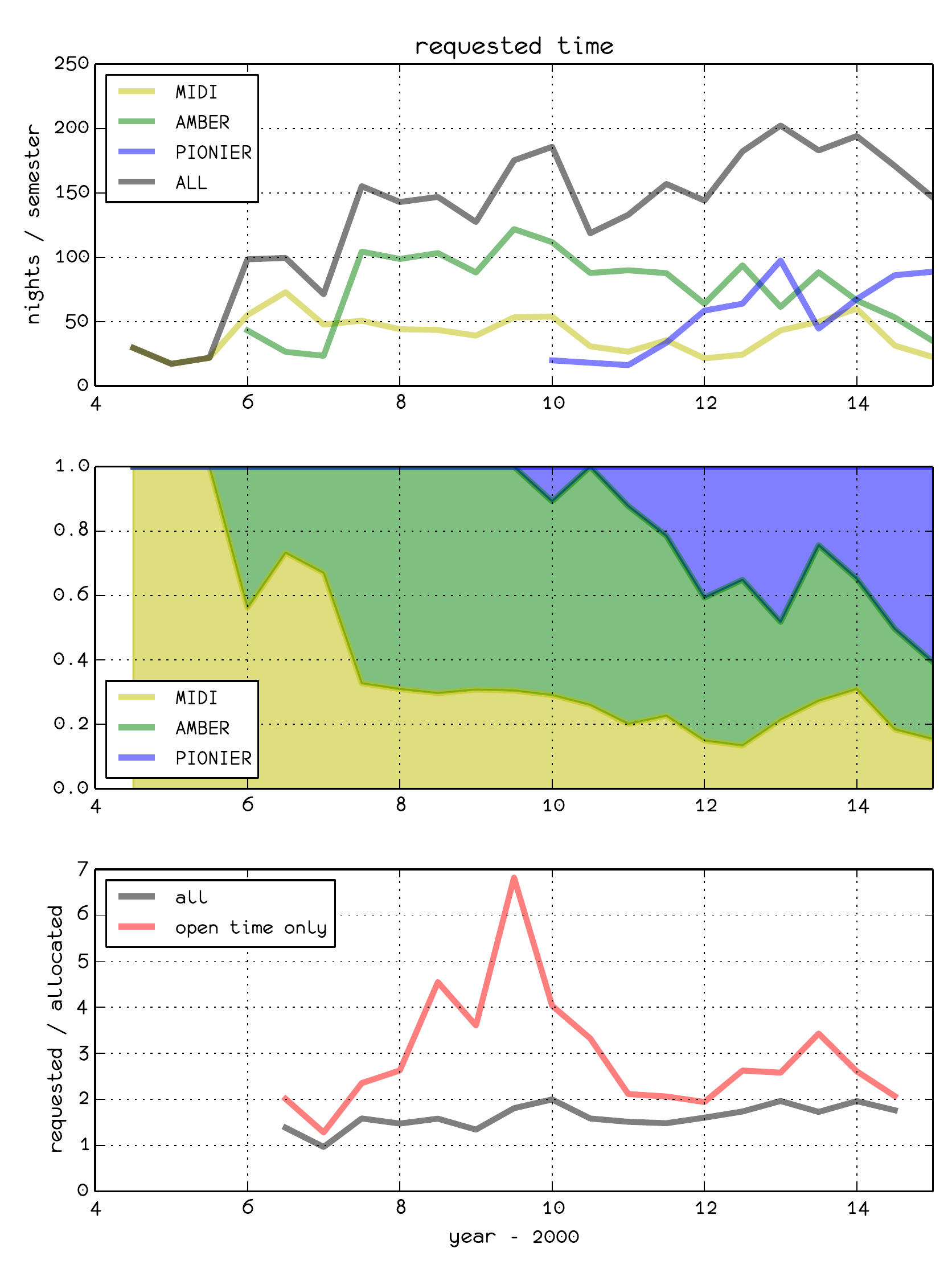}
   \includegraphics[height=9.1cm]{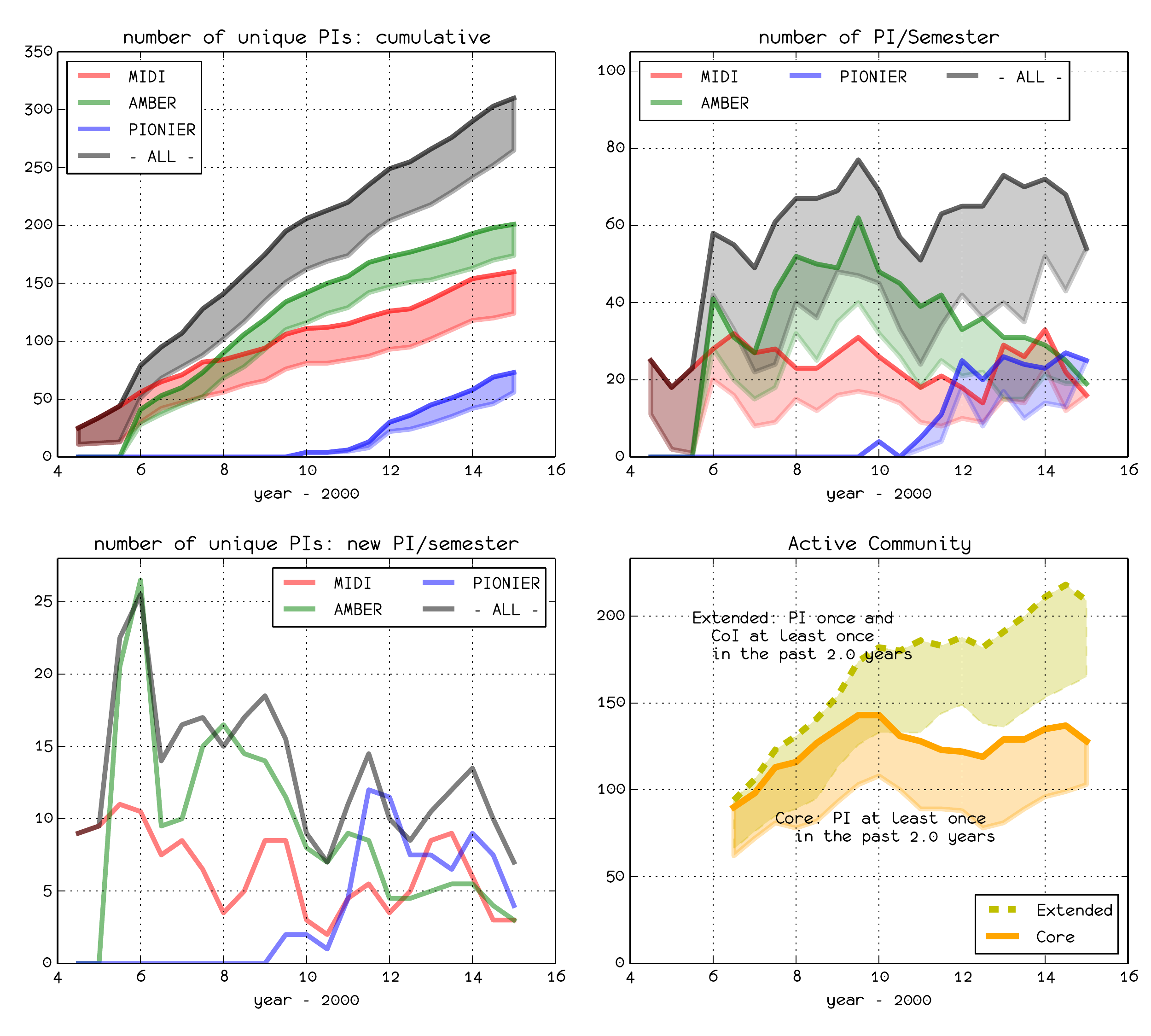}
   \end{center}
   \caption[Community]{\it \small \label{fig:Community} Requested time (left most panels) and Users Community of astronomers requesting and obtaining time (center and rightmost panels) at VLTI in the past 10 years: for each plot with thick curves, the upper curve is for the community of requesters, and the lower plot the community of people obtaining time.}
   \end{figure} 

The user community of the VLTI can be studied using the submitted observing proposal received by ESO every semesters. The two angles we propose to look at are: 1) the number of requested time per semester and per instruments 2) the evolution of the PI (principal investigators) community.

The requested time has been stable over the years, between 150 and 200 nights per semester (Fig.~\ref{fig:Community}, upper left panel), even though the requests are a bit different between the 3 offered instruments: MIDI has always been stable, around 50 nights/semester, decreasing slightly to 30 but gaining interest again with the introduction of a new fringe tracking mode\cite{midifsu}; AMBER peaked at 120 nights/semester in 2009, and has decreased to 40 nights/semester; PIONIER, which is a visitor instrument, is now being requested for 80 nights/semester. Currently, PIONIER is the most requested instrument, with 60\% of the requested time at VLTI (Fig.~\ref{fig:Community}, center left panel). The overall time pressure on VLTI is around two. However, in these numbers, the Guaranteed Time Observing (GTO) is not taken into account. Considering only the open time, the pressure varied a lot more in the past, peaking close to 7 in 2009, when the AMBER consortium used the bulk of its guaranteed time. Since then, the pressure is around 3 (Fig.~\ref{fig:Community}, lower left panel).

\subsection{PRINCIPAL INVESTIGATORS COMMUNITY}

Each Semester, about 60 unique PIs request time at VLTI, and many more Co-Is (see Fig.~\ref{fig:Community}, upper right panel). Among these PIs, about 10 are completely new to VLTI. Over the past 10 years, more than 300 unique PIs have submitted observing proposals (see Fig.~\ref{fig:Community}, upper middle panel). The different instruments have different behaviors in appearance (see Fig.~\ref{fig:Community}, lower middle panel), but they actually follow the same pattern: The number of new PIs peaks soon after the introduction of the instrument and slowly decrease after. AMBER is a good example of that, and PIONIER is as well, even if its short career does not allow us to conclude firmly. MIDI seems different, but the successful introduction of MIDI+FSU\cite{midifsu, midifsuMueller} in 2012 led to a regain of interest for the instrument soon after.

If one considers the PIs obtaining time, about 40 out of 60 PIs obtain time each semester. The total of PIs who obtained time is more than 260 PIs, which is a good number: almost 90\% of the PIs who ever applied for time obtained some. If one defines the Core Community as the set of PIs who have submitted at least one proposal in the past 2 years, the VLTI has a core community of about 130 unique requester PIs, and about 80 of people obtaining time (see Fig.~\ref{fig:Community}, lower right panel). The size of the Core Community is changing and seems to be related to the availability/popularity of new instruments: the Core Community started to shrink in 2009, corresponding to the peak of interest for AMBER, and started to grow again once PIONIER was popularized in 2012\footnote{This is also confirmed by looking at other communities of observation technique at the VLT}. This is also seen using an Extended Community definition (PIs once in the past and Co-Is in the last 2 years), which seems to be more prone to grow in time. The Extended Community of requesters is currently of the order of 200 people, and 150 people if one counts only Co-Is obtaining time in the past 2 years.

For comparison, the Core Community of Adaptive Optics imaging at VLT (using the SINFONI and NACO instruments) is about 2.5 times bigger than the VLTI community, and shows the same kind of time evolution. For example, the growth, in term of new PIs who never applied before, show the same variations: it peaks soon after the introduction of new instruments and slowly decreases after that. In other words, communities grow when new instruments or new capabilities are offered (such as PIONIER in $\sim$2011 or MIDI+FSUA in $\sim$2013).
 
\subsection{NETWORK AND RELATIONS}

\begin{figure}
   \begin{center}
      \includegraphics[width=18.5cm, trim=240 0 100 0, clip=true]{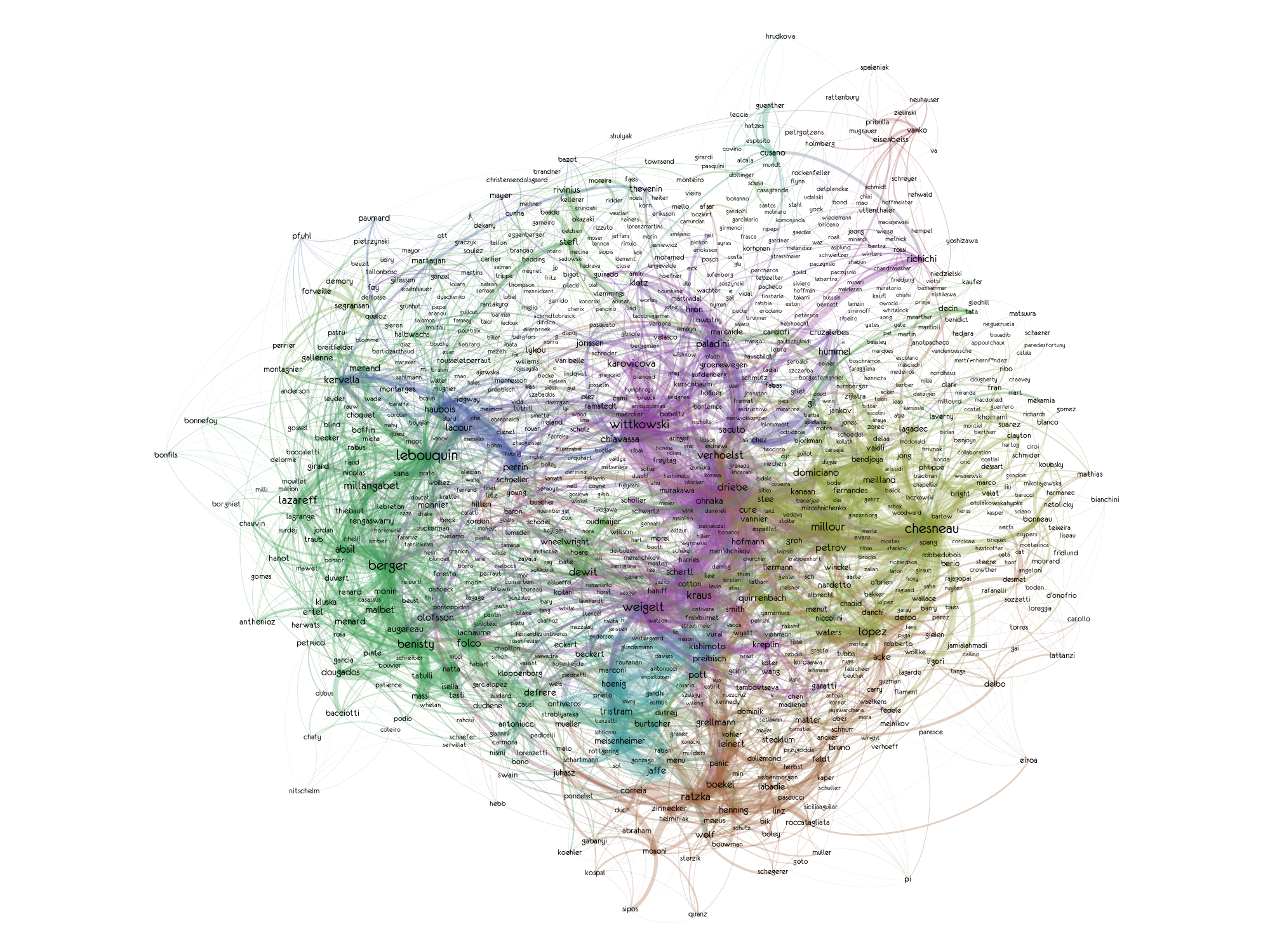}
   \end{center}
   \caption[Network]{\it \small \label{fig:Network} Network of collaborations ($\sim$1000 PIs and Co-Is in submitted proposals) in the VLTI community for the period 2003-2014, for instruments AMBER, MIDI and PIONIER. The size of the dots and labels are proportional to the number of connections, the colors represents the different groups of collaborators using a clustering algorithm\cite{clustering}. This figure was generated using Gephi\cite{gephi} (\texttt{www.gephi.org})}
 \end{figure} 

Another way to look at the community is to plot the network of relations derived from the proposals list of principal and co-investigators. A way to do so is to collect the PI-CoI connection and plot a network using a force-directed graph drawing algorithm ("Force Atlas", from Gephi\cite{gephi}). Then, using a clustering algorithm\cite{clustering}, one can find the different group of collaborators. Doing so, a few patterns show up (see Fig.~\ref{fig:Network}):
\begin{itemize}
\item The original MIDI, AMBER and PIONIER groups are strong clusters: Henning, Jaffe, Ratzka etc. for MIDI; Petrov, Millour, Weigelt etc. for AMBER; Lebouquin, Berger, Lazareff, Millan-Gabet etc for PIONIER;
\item the clustering seems not only dominated by the instruments, but also the institutes of the persons: \textbf{Nice} (Petrov, Millour, Chesneau, Domiciano de Souza, Meilland, Nardetto etc), \textbf{Bonn} (Weigelt, Driebe, Kraus, Ohnaka, etc.), \textbf{Heidelberg} (Henning, von Boeckel, Ratzka, Wolf, etc.), \textbf{Grenoble} (Lebouquin, Benisty, Malbet, Lazareff, etc.), \textbf{Paris} (Kervella, Haubois, Perrin, Lacour, etc.), \textbf{Vienna} (Sacuto, Paladini, Hron, etc.), \textbf{ESO-Garching} (Wittkowski, Karovicova, Richichi, etc.), \textbf{ESO-Chile} (Rivinius, Stefl, de Wit, Martayan, Gallenne, M\'erand, Boffin, etc.)
\item MATISSE is well represented because of their early involvements in MIDI and AMBER: Lopez (PI), Henning, Weigelt, Jaffe, Wolf, etc.
\item GRAVITY is at the periphery and dominated by the Paris contribution: Perrin, Kervella, Lacour, Haubois, Paumard. The PI, F. Eisenhauer, is not very visible in the current VTLI network, which is not a surprising since his group work was not in interferometry.
\end{itemize}

\subsection{PUBLICATIONS}

\begin{figure}
   \begin{center}
   \includegraphics[width=16.5cm]{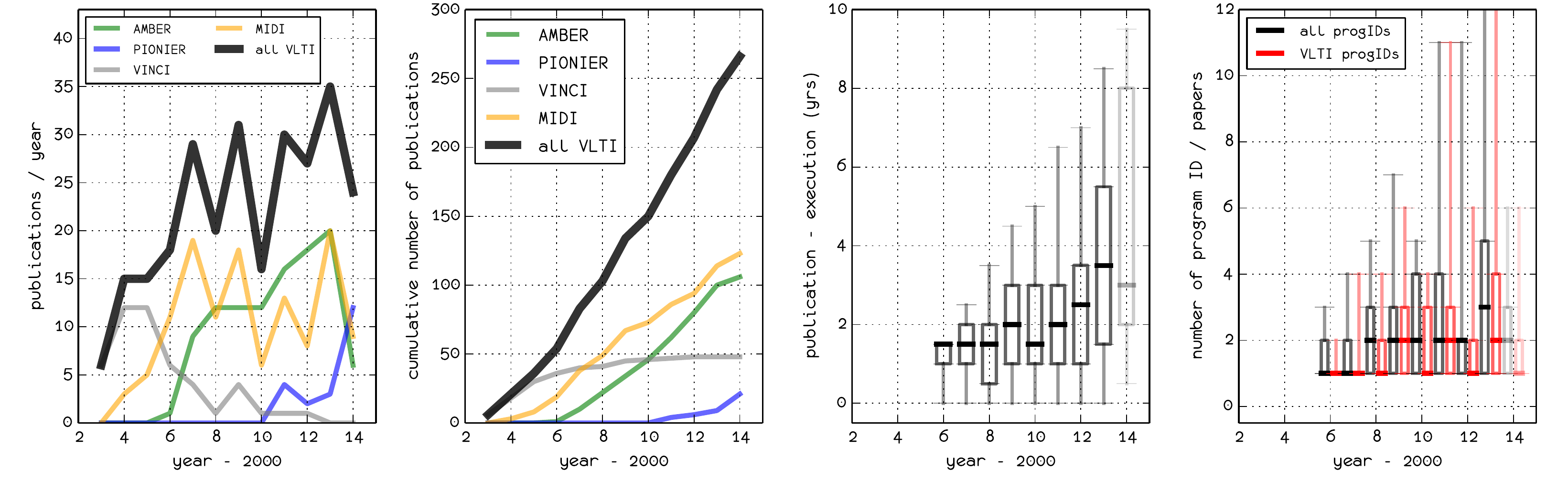}
   \end{center}
   \caption[Publications]{\it \small \label{fig:Publications} \textbf{Bibliometric of the VLTI}, from left to right: 1) Publications rate; 2) total number of papers; 3) difference, in years, between publication and observations 4) number of ESO program ID reported in the paper, for all instruments (black) and VLTI only (red). The abscise is the year of publication. For the two rightmost panels, for each year, the symbol show the median (thick line), the first and third quartiles (box) and the minimum and maximum of the distribution (line with ticks). Note that a paper can be counted for various instruments, if it uses data from these instruments, but it will be, however, counted only once in the "all VLTI" category. Note also that the numbers for 2014 are extrapolated based on the first 4 months of that year.} 
\end{figure} 

The ESO online bibliographic tool\cite{2012SPIE.8448E..21G} (\texttt{http://telbib.eso.org}) allows tracking the usage, in term of refereed science publication, of VLTI. Fig.~\ref{fig:Publications} shows the evolution, over the last decade, of the publication rate, as well as the cumulative number of papers. The publication rate seems stable since 2007, corresponding to the analysis that the active community of users has also been stable in the past decade (see Fig.~\ref{fig:Community}). If we compare the publication rate and total number of publication of the VLTI community to the VLT-AO community (NACO and SINFONI instruments) we find the same scaling: in absolute number, the VLTI community publishes $\sim$2.5 less than the VLT-AO community, which is also the ratio of the size of the communities. In other words, VLTI and VLT-AO imaging have similar publication rate compare to their communities' respective sizes.

It is also interesting to consider the year of publication compared to the year(s) the observing program(s): we plotted such information on the third panel form the left of Fig.~\ref{fig:Publications}. The median time to publish a paper after the data were taken seems to be of the order of 2 years, increasing a little bit with time (from $\sim$1.5~year in 2006 to $\sim$3~years in 2013). The reason for this increase is hard to identify without reading all the papers in details: it could be either due to more complex data to analyze or the inclusion of archival data in order to complete the result of an analysis. 

A hint of answer is found in the number of ESO program ID (i.e. number of observing proposals) reported per paper: it is increasing with time. This is a testimony that the authors of papers combine more and more various types of observations, either because they are interferometrists using other technics or, conversely, they come from another background and use interferometry to complement their observations. This analysis is limited in its scope, since it only considers ESO observations programs, but still, the signs are pointing toward VLTI being more and more used in combination with other techniques.

\subsection{THOUGHTS ON THE VLTI COMMUNITY}

The VLTI has a relatively small community: overall, 10\% of the VLT PIs over the past 10 years applied for VLTI time at one point or the other. Compared to other techniques, it is $\sim$2.5 smaller than the VLT-AO imaging community $\sim$5 times smaller than the VLT-slit spectroscopy community. The VLTI community does not seem intrinsically different from others: the publication rate and growth (as well as other indicators) scale with its size as expected. 

The community has grown around instruments and power users who drew more PIs into the pool of submitters. New instruments and new modes are the trigger for rapid growth of the community, and the upcoming upgrades and new instruments surely will have the same effect. We briefly present in the next section some of these improvements. Most of these aspects are covered in specific papers in these proceedings.

\section{NEAR FUTURE OF VLTI (2015-2017)}

The VLTI facility has to keep evolving to maintain the scientific 
relevance. The upcoming evolutions of the VLTI include the commissioning of:
\begin{itemize}
\item The star separators for the Auxiliary and Unit Telescopes (expected completion in early 2015).
\item The near-infrared spectro-interferometric and astrometric 4T instrument GRAVITY ($\sim$2015)
\item The mid-infrared spectro-interferometric instrument MATISSE ($\sim$2016)
\item The adaptive optics facilities for the Auxiliary Telescopes NAOMI ($\sim$2017) 
\end{itemize}




\subsection{STAR SEPARATORS}

\begin{figure}
   \begin{center}
   \includegraphics[width=12cm]{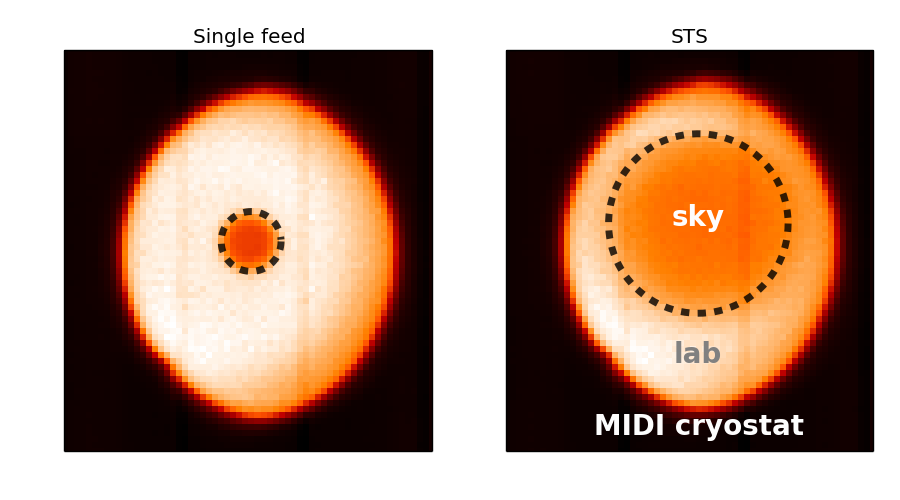}
   \end{center}
   \caption[STSfov]{\it \small \label{fig:STSfov} Comparison of the field-of-view in the MIDI instrument for the Single Feed (left) and for one field of the Star Separator (STS) on the ATs. The inner most darker disk is the on-sky field-of-view, delimited by the dashed gray line; the almond shape limit between the dark and clear area is the MIDI field stop; between the two is the emission from the VLTI laboratory, which appear brighter than the sky. The field-of-view of the single feed is $\sim$2" in diameter, while $\sim$6" for the STS.}
\end{figure} 

The star separators (STS) were first envisioned to be used for PRIMA astrometric facility\cite{primaSahlmann, prima}~. In the case of the ATs, the star separator is a Relay Optic System (ROS) distinct from the single feed one. The STS have the following characteristics: 

\begin{center}       
\begin{tabular}{|p{5cm}|p{5cm}|p{5cm}|} 
\hline
 & \bf Single-Feed &  \bf Dual-Feed (STS) \\
\hline
AT field-of-view (see also Fig.~\ref{fig:STSfov}) & 1" to 4" depending on the AT station and DL position& full field of view ($\sim$5"), comparable to UT field-of-view\\
\hline
maximum DL-VCM pressure & $\sim$5 bar & $\sim$1.8 bar\\
\hline
fast pupil actuator & no & yes \\
\hline
differential transmission STS/SF & \centering --- & $\sim$0.50 (J band), $\sim$0.65 (H, K bands) and $\sim$0.70 (N band) \\
\hline
\end{tabular}
\end{center}

Based on the characteristics, it has been decided to operate in the future only with STS ROS, in order to improve the performances (field of view and pupil relay), and maintain the number of systems to maintain (4). This is at the cost of a loss in transmission. In particular, the field of view increase and lowered pressure of the Variable Curvature Mirror (VCM) should bring performance and operational gains. 

\subsection{GRAVITY and MATISSE}

GRAVITY\cite{gravity} and MATISSE\cite{matisse} are the second generation instruments in development for the VLTI. GRAVITY has two modes of operations: it can be used as a 4T spectro-interferometer in K band, including a fringe tracker, or as a 4T astrometric instrument. 

In terms of changes to the infrastructure, the following can be noted:
\begin{itemize}
\item GRAVITY will take the place of the Visitor Focus, occupied currently by PIONIER. The visitor focus will be moved, and PIONIER will be moved to the new location;
\item GRAVITY-astrometry will use a laser metrology to measure the optical path from its beam combiners to each telescope's entrance pupil. This metrology will require installation of additional hardware on the ATs and UTs secondary mirror spider;
\item GRAVITY will come with infrared wavefronts sensors for the UT adaptive optics, MACAO, which currently sense the wavefront in the visible;
\item MATISSE will take the place of MIDI in the VLTI laboratory. MIDI will be removed at the time GRAVITY arrives, in order to minimize the work;
\item All the changes require adaptation of the VLTI software infrastructure, in particular the Interferometer Supervisor Software (ISS).
\end{itemize}

\subsection{NAOMI}
\label{sec:naomi}
NAOMI is the future adaptive optics system for the ATs, a low-order Shack-Hartmann system operating in the visible in place of the current tip-tilt correction. The goal is to provide improved coupling in the combiners (AMBER, PIONIER, GRAVITY and MATISSE), in order to improve the signal-to-noise ratio, especially when the conditions are poor. Thanks to NAOMI, we expect not only a good input wavefront, but we expect it more often, as we have shown that weather losses are "built-in" at VLTI (see Fig.~\ref{fig:VLTIusage} and \ref{fig:ASMstats}): we can hope to be operational 20\% more often (for seeing in the $\sim$1.0"-1.5" regime). A dedicated paper in this conference describes the project in more details\cite{naomi}.

\section{CONCLUSIONS}

VLTI on at the verge of big changes: the facility is going to evolve and new instruments are coming. This could be felt as unfortunate, because the VLTI as it is now is operationally more effective than it has ever been: according to our metric (see Sec.~\ref{sec:perfo}) it improved by a factor of 12. VLTI also serves a growing number of unique PIs (more than 250 as of 2014, about 10\% of the VLT PIs) and a stable active community (about 100 PIs). On the other hand, growth of the community (new PIs of observing proposals, hence new science topics) is provided by new modes (such as MIDI+FSUA) and/or new instruments (such as PIONIER). 

Implementing the changes required for the second generation instruments will be challenging, but the result will be improved performances. The second challenge will be to maintain the operational efficiency we have reached in the past few years. 


\acknowledgments     

This publication reflects on the work not only of the past two years (in the form of the list of the co-authors representing the active team covering this period), but covers also the efforts of the collective work in the past 12 years, which is not entirely covered by the co-authors list. We wish here to acknowledge this collective work of all the persons, from ESO and in the community, associated at one point or the other with the development and operations of the VLTI who are no listed in the co-authors' list. 
 

\bibliography{vltibib}   
\bibliographystyle{spiebib}   

\end{document}